\documentclass[aps,prb,preprint,superscriptaddress,colorlinks=true,linkcolor=black,urlcolor=black,citecolor=black,anchorcolor=black,colorlinks,bookmarks=false,citecolor=blue,linkcolor=black,urlcolor=blue]{revtex4-1}

\usepackage{ulem}
\usepackage{bm,color,amsmath,amssymb,latexsym,graphicx,psfrag,accents,float}
\usepackage{amscd,dsfont,wasysym,mathrsfs,mathtools}
\usepackage{multirow}
\usepackage{dcolumn}
\usepackage{xcolor}
\usepackage{comment}
\usepackage{booktabs} 
\usepackage{upgreek}
\usepackage[utf8]{inputenc}
\usepackage{braket}
\usepackage{times}
\usepackage{amsfonts}
\usepackage{mathrsfs}
\usepackage{graphicx}
\usepackage{dcolumn}
\usepackage{bm}
\usepackage{color}

\usepackage[colorlinks,bookmarks=false,citecolor=blue,linkcolor=red,urlcolor=blue]{hyperref}
\usepackage{amsmath}
\usepackage{ulem}
\begin{document}
\renewcommand{\theequation}{S\arabic{equation}}
	\newcommand{\beginMethods}{%
		\setcounter{table}{0}
		\renewcommand{\thetable}{S\arabic{table}}%
		\setcounter{figure}{0}
		\renewcommand{\thefigure}{S\arabic{figure}}%
	}
\newcommand{\bk}{{\bf k}}
\newcommand{\bB}{{\bf B}}
\newcommand{\bv}{{\bf v}}

\title{Switchable chiral transport in charge-ordered Kagome metal CsV$_3$Sb$_5$}
\author{Chunyu Guo${}^{\dagger}$}\affiliation{Laboratory of Quantum Materials (QMAT), Institute of Materials (IMX),\'{E}cole Polytechnique F\'{e}d\'{e}rale de Lausanne (EPFL), Lausanne, Switzerland}
\affiliation{Max Planck Institute for the Structure and Dynamics of Matter, Hamburg, Germany}
\author{Carsten Putzke${}^{}$}
\affiliation{Max Planck Institute for the Structure and Dynamics of Matter, Hamburg, Germany}
\author{Sofia Konyzheva}\affiliation{Laboratory of Quantum Materials (QMAT), Institute of Materials (IMX),\'{E}cole Polytechnique F\'{e}d\'{e}rale de Lausanne (EPFL), Lausanne, Switzerland}
\author{Xiangwei Huang${}^{}$}\affiliation{Laboratory of Quantum Materials (QMAT), Institute of Materials (IMX),\'{E}cole Polytechnique F\'{e}d\'{e}rale de Lausanne (EPFL), Lausanne, Switzerland}
\author{Martin Gutierrez-Amigo}\affiliation{Centro de Física de Materiales (CSIC-UPV/EHU), Donostia-San Sebastian, Spain}
\affiliation{Department of Physics, University of the Basque Country (UPV/EHU), Bilbao, Spain}
\author{Ion Errea}\affiliation{Centro de Física de Materiales (CSIC-UPV/EHU), Donostia-San Sebastian, Spain}
\affiliation{Donostia International Physics Center, Donostia-San Sebastian, Spain}
\affiliation{Fisika Aplikatua Saila, Gipuzkoako Ingeniaritza Eskola, University of the Basque Country (UPV/EHU), Donostia-San Sebastian, Spain}
\author{Dong Chen}\affiliation{Max Planck Institute for Chemical Physics of Solids, Dresden, Germany}\affiliation{College of Physics, Qingdao University, Qingdao, China}
\author{Maia G. Vergniory}\affiliation{Donostia International Physics Center, Donostia-San Sebastian, Spain}
\affiliation{Max Planck Institute for Chemical Physics of Solids, Dresden, Germany}
\author{Claudia Felser}\affiliation{Max Planck Institute for Chemical Physics of Solids, Dresden, Germany}
\author{Mark H. Fischer${}^{\dagger}$}\affiliation{Department of Physics, University of Zürich, Zürich, Switzerland}
\author{Titus Neupert${}^{\dagger}$}\affiliation{Department of Physics, University of Zürich, Zürich, Switzerland}
\author{Philip J. W. Moll${}^{\dagger}$}\affiliation{Laboratory of Quantum Materials (QMAT), Institute of Materials (IMX),\'{E}cole Polytechnique F\'{e}d\'{e}rale de Lausanne (EPFL), Lausanne, Switzerland}
\affiliation{Max Planck Institute for the Structure and Dynamics of Matter, Hamburg, Germany}

\date{\today}
\maketitle
\normalsize{$^\dagger$Corresponding authors: chunyu.guo@mpsd.mpg.de(C.G.); mark.fischer@uzh.ch(M.H.F.); titus.neupert@uzh.ch(T.N.); philip.moll@mpsd.mpg.de(P.J.W.M.).}

\section*{SUMMARY PARAGRAPH}
When electric conductors differ from their mirror image, unusual chiral transport coefficients appear that are forbidden in achiral metals, such as a non-linear electric response known as electronic magneto-chiral anisotropy (eMChA)\cite{TeChiral,BiHelix,carbon,TTlO4,CrNbSB,MnSi}. While chiral transport signatures are by symmetry allowed in many conductors without a center of inversion,  it reaches appreciable levels only in rare cases when an exceptionally strong chiral coupling to the itinerant electrons is present. So far, observations of chiral transport have been limited to materials in which the atomic positions strongly break mirror symmetries. Here, we report chiral transport in the centro-symmetric layered Kagome metal CsV$_3$Sb$_5$, observed via second harmonic generation under in-plane magnetic field. The eMChA signal becomes significant only at temperatures below $T'\sim$ 35 K, deep within the charge-ordered state of CsV$_3$Sb$_5$ ($T_{\mathrm{CDW}}\sim$ 94 K). This temperature dependence reveals a direct correspondence between electronic chirality, unidirectional charge order\cite{zhao2021cascade}, and spontaneous time-reversal-symmetry breaking due to putative orbital loop currents\cite{Mielke2022,yu2021evidence,Nie2022}. We show that the chirality is set by the out-of-plane field component and that a transition from left- to right-handed transport can be induced by changing the field sign. CsV$_3$Sb$_5$ is the first material in which strong chiral transport can be controlled and switched by small magnetic-field changes, in stark contrast to structurally chiral materials -- a prerequisite for their applications in chiral electronics.

\section*{MAIN TEXT}
The role symmetries play in determining the properties of matter can hardly be overstated. Two opposite extremes are particularly interesting in crystalline solids: Higher symmetries constrain emergent degrees of freedom to mimic free particles — creating, for instance, massless Dirac or Weyl fermions that recover at low energies almost the full Lorentz group of free space\cite{RMP_topo,RMP_topo0,RMP_topo1,RMP_topo2}. A second angle is to study low symmetry systems with novel responses. Among these, asymmetric systems characterized as \textit{chiral} play a special role across biology, chemistry, and physics\cite{ChChem,ChSpin}. Crystals are structurally chiral if they possess no mirror, inversion or roto-inversion symmetry, giving rise to left- and right-handed enantiomers. This chirality can be imprinted on the crystals' emergent excitations, which are then also characterized by a definite handedness. The interaction between structural chirality and the breaking of time-reversal symmetry (TRS) is of particular interest, as it links the static chirality to temporal processes such as growth, catalysis, and wave propagation\cite{ChSky}. Response functions that jointly arise due to chirality and TRS breaking are called magneto-chiral anisotropies\cite{MCHA}. Specifically, in metals one studies the electronic magneto-chiral anisotropy (eMChA), which opens up possibilities to detect, manipulate and utilize chiral properties in electronics\cite{TeChiral,BiHelix,carbon,TTlO4,CrNbSB,MnSi}.

\begin{figure}
	\centering
	\includegraphics[width=0.86\columnwidth]{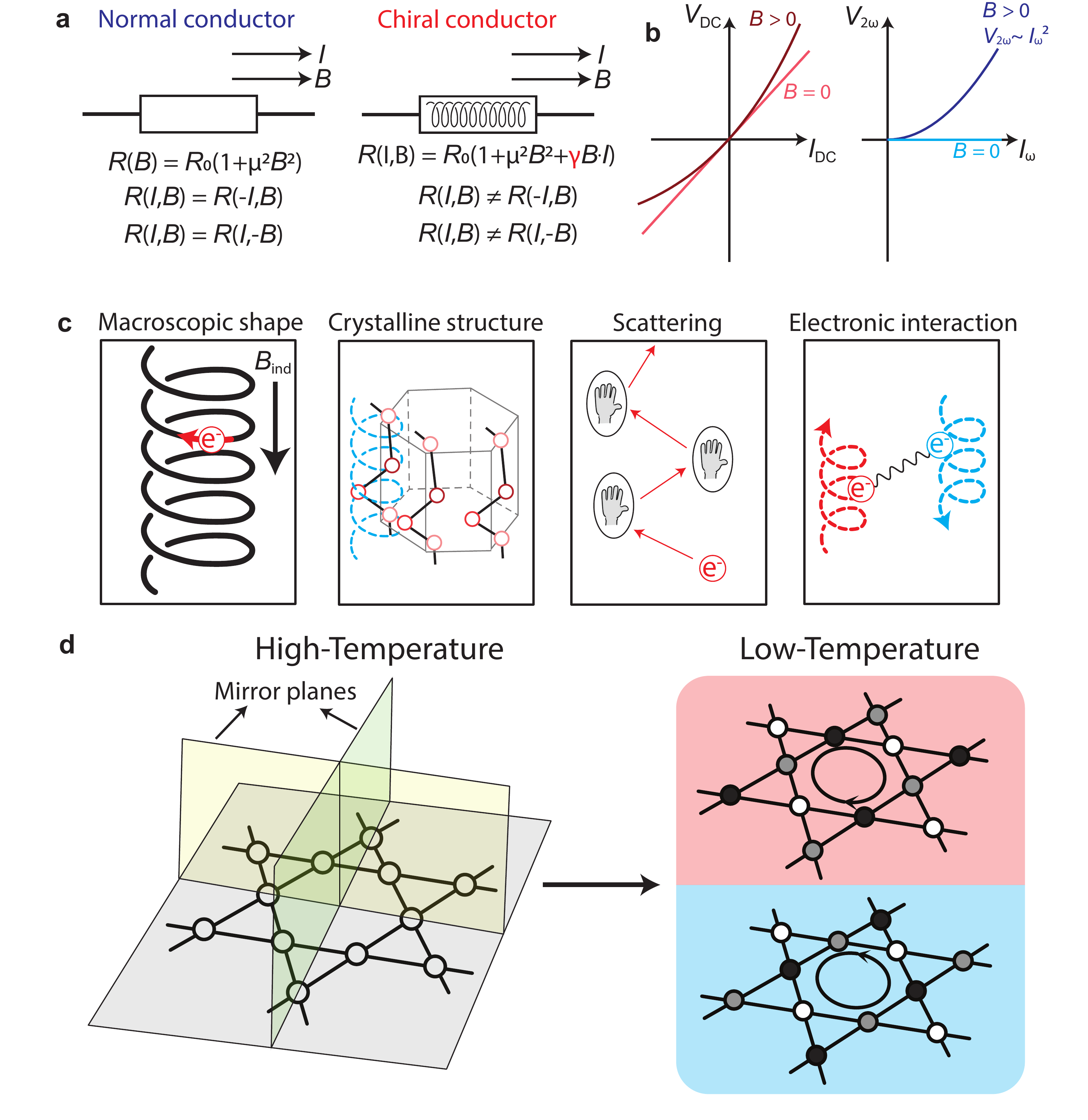}
		\caption{\textbf{Electronic magneto-chiral anisotropy (eMChA) and spontaneous symmetry breaking in CsV$_3$Sb$_5$.} (a) Illustration of electrical resistance of normal and chiral conductors within the low-frequency limit. (b) I(V) curve for a chiral conductor. In a DC measurement (left), the measured voltage displays a non-linear current dependence with magnetic field applied. In the AC case (right), the field-induced second harmonic voltage $V_{2\omega}$ depends quadratically on AC current $I_{\omega}$. (c) Different mechanisms for  electronic magneto-chiral anisotropy. The blue dashed line in crystalline structure case represents the notation of helical atomic chain. For the case of scattering the encircled hands represent the scattering centers with particular chirality. (d) The crystal structure of CsV$_3$Sb$_5$ preserves all mirror symmetries at high temperatures and only spontaneous symmetry breaking at low temperatures allow for a finite eMChA in a symmetric microstructure. }
	\label{eMChA}
\end{figure}

 \textcolor{black}{eMChA} usually refers to a change in resistance $R$ due to an applied current $\boldsymbol{I}$ and external magnetic field $\boldsymbol{B}$ that is conventionally expressed as $R(\boldsymbol{B},\boldsymbol{I}) = R_0 (1+\mu^2 \boldsymbol{B}^2 + \gamma^\pm \boldsymbol{B}\cdot \boldsymbol{I})$\cite{TeChiral}, see Fig.~\ref{eMChA}. Time-reversal symmetry in non-magnetic metals enforces a magnetoresistance even in field, which usually takes the semi-classical form $\mu^2 \boldsymbol{B}^2$ with $\mu$ the mobility. The scalar product $\boldsymbol{B}\cdot \boldsymbol{I}$ is only allowed in chiral crystals without space-reflection symmetries and hence eMChA appears. Its strength is described by the coupling constant $\gamma^\pm$ which takes opposite sign for the two enantiomers and is tensorial in anisotropic conductors.  eMChA is most commonly detected by the associated second-harmonic voltage generation under low-frequency AC currents,  $ 4V_{2\omega}/V_{\omega} = \Delta R / R$, where $\Delta R = R(\boldsymbol{B},\boldsymbol{I}) - R(\boldsymbol{B},-\boldsymbol{I})$ denotes the odd-in-current and $R =R(\boldsymbol{B},\boldsymbol{I}) + R(\boldsymbol{B},-\boldsymbol{I})$ denotes the even-in-current contribution to the resistivity. 

To display eMChA, a conductor must break inversion symmetry, which can occur as a weak effect in any metal when its macroscopic shape is chiral\cite{BiHelix,carbon}, for example in a coil (Fig.~\ref{eMChA}). Alternatively, materials with chiral crystal structure\cite{TeChiral,TTlO4} generally show eMChA in any conductor shape (We note that the ``chiral electronic structure'' of the symmetry-broken phase mentioned here does not necessarily have to have the symmetries of a chiral space group. In a layered, quasi-2D compound one refers to a structure as ``chiral'' when the in-plane mirror symmetries are broken, while the $M_z$ mirror symmetry might still be intact. However, this lack of in-plane mirror symmetries is enough to enable the observation of eMChA in the geometry of our measurements.). eMChA expresses an imbalance between scattering processes of different handedness, which can either occur from the intrinsic handedness of the carriers in chiral crystals, or extrinsically from chiral defects as in plastically twisted conductors. When electronic interactions form ordered phases within chiral materials, as for example in chiral magnets, eMChA can be further amplified via scattering off, e.g., an emergent chiral spin texture\cite{CrNbSB,MnSi}. 






In this work, we demonstrate eMChA in a rectangular bar of CsV$_3$Sb$_5$, a layered metal in which vanadium atoms form Kagome nets. In this system, a cascade of correlated symmetry-breaking electronic phases emerges at low temperatures\cite{zhao2021cascade,liang2021three,Xiang2021,Kang2022,Nie2022,TitusAdd}, including a charge-density-wave (CDW) state below $T_{\mathrm{CDW}} \approx 94 $ K and superconductivity below $T_{c} \approx 2.5 $ K \cite{ortiz2019new,ortiz2020cs,yu2021concurrence,mu2021s,liang2021three,Neupert2022}. Experimental evidence mounts for a further transition within the charge-ordered phase at $T'\approx 35$ K, accompanied by an additional 4a$_0$ unidirectional ordering vector\cite{zhao2021cascade} and time-reversal symmetry breaking\cite{Mielke2022,yu2021evidence,Nie2022}. The sudden onset of unexpectedly strong chiral transport at $T'$ is our main observation. Crucially, this system is centro-symmetric at high temperatures, yet the relevant mirror symmetries are spontaneously broken by correlated phases of the itinerant carriers (Fig. \ref{eMChA}). Reversible chirality of the electronic structure within the CDW phase has been observed in scanning tunneling microscopy (STM) experiments~\cite{Kchiral}. Note that the accompanying crystal distortion is so weak that the low-temperature crystal structure remains actively debated\cite{ortiz2020cs,zhao2021cascade,Luo2022,XrayCD}. In contrast to structurally chiral crystals that strongly differ from their mirror image, here the differences between the enantiomers are subtle and test the limits of experimental resolution. Hence the observation of eMChA itself in this compound points to its novel origin. As a consequence, the material's chirality itself can be switched which leads to the field-switchable chiral transport in CsV$_3$Sb$_5$.


 To truly obtain a symmetry lowering from spontaneous symmetry breaking, it is critical to avoid any accidental strain fields that may break the symmetry explicitly. To do so, we decouple the crystalline bar mechanically as much as possible from its supporting substrate\cite{huang2022mixed} (Fig. \ref{Cs135}a). This structure is mechanically supported by the gold-coated SiN$_x$ membrane-based (200~nm thick) springs, the differential thermal contraction strain is estimated to be less than 20 bar. Any signatures of chiral transport vanish in a reference experiment with even modest strain fields caused by stiff substrate coupling (see Methods), evidencing a strong coupling between the charge order and lattice distortions, which is not surprising in CDW systems\cite{CDW_RMP}. This provides a natural explanation for the opposing STM experiments\cite{li2022rotation,Kchiral}.



\subsection*{Observation of eMChA in CsV$_3$Sb$_5$}
\begin{figure}
	\centering
	\includegraphics[width=0.86\columnwidth]{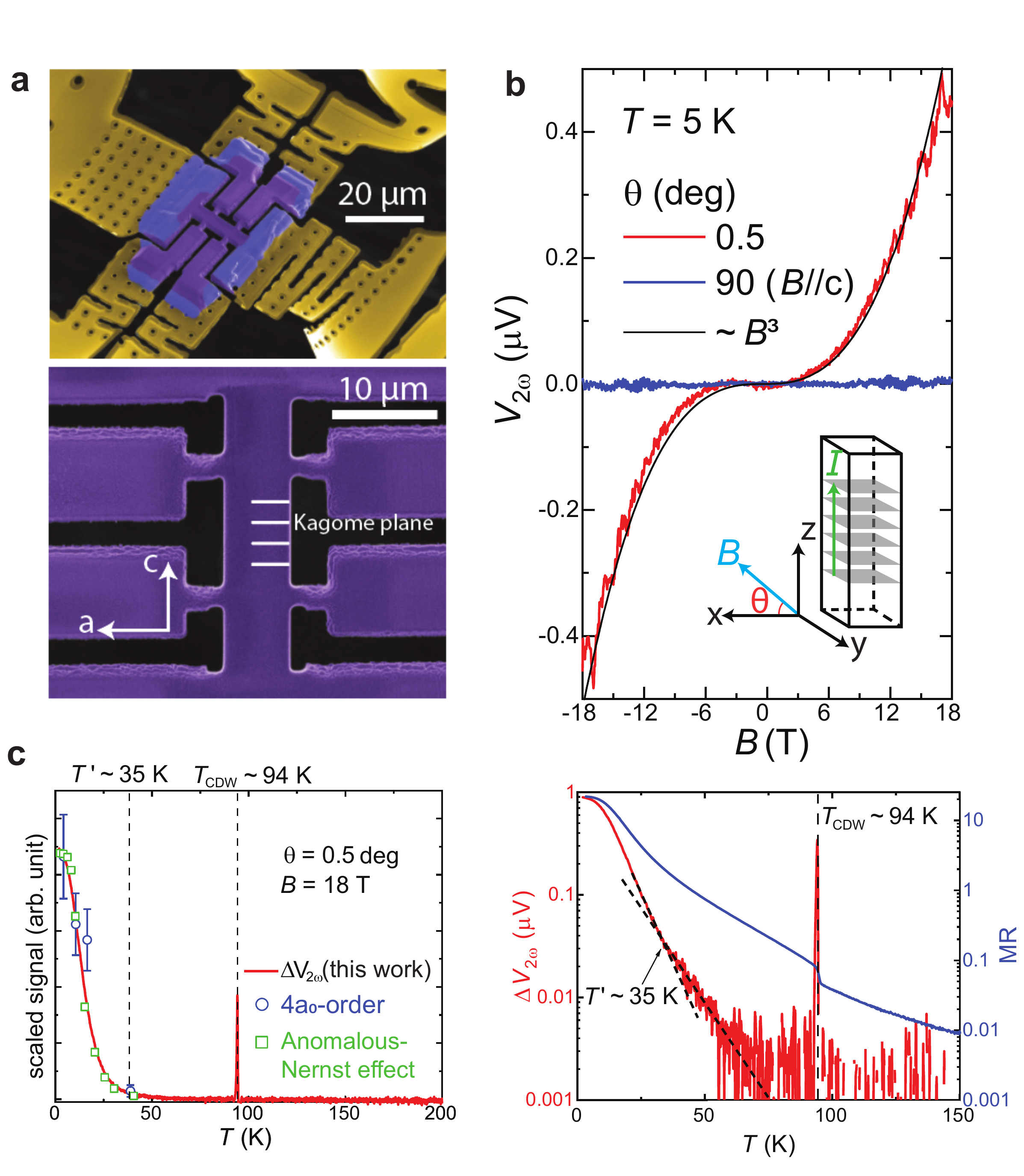}
		\caption{\textbf{Field- and temperature-dependence of eMChA.} (a) Low-strain microstructure fabricated by FIB. (b) Field dependence of second harmonic voltage with current applied along z(c)-axis. The signal becomes sizable when magnetic field is applied approximately in the plane. The inset sketches the transport bar. (c) The red continuous line in the left panel represents temperature dependence of $\Delta V_{2\omega} = V_{2\omega}(18\,\mathrm{T})-V_{2\omega}(-18\,\mathrm{T})$ with magnetic field applied approximately in the plane. The blue open circles show the FT intensity of the wavevector corresponding to the 4a$_0$ unidirectional charge order phase adopted from Ref.\cite{zhao2021cascade} and the green squares represent the anomalous Nernst effect reported in Ref.\cite{SAxy}. The right panel displays log-scale temperature dependence of $\Delta V_{2\omega}$ and magnetoresistance ratio MR = ($\rho_c$(18 T)-$\rho_c$(0))/$\rho_c$(0). }
	\label{Cs135}
\end{figure}

Our main observation is the appearance of a sizable second harmonic response, $V_{2\omega}$, to a low-frequency transport current (7 Hz) which distinctly evidences the diode-like behavior due to chiral transport within the charge-ordered state at low temperatures (Fig. \ref{Cs135}). First, we discuss out-of-plane currents under an approximately in-plane magnetic field which is purposely misaligned by 0.5° with respect to the Kagome planes. At zero field and $T$ = 5 K just above the superconducting transition, no second harmonic is observed, yet the signal quickly grows with increasing magnetic field. Its field-dependence is well described by $V_{2\omega}\propto B^3$ up to 18 T, the highest fields accessible to the experiment. This strikingly departs from the behavior of structurally chiral materials such as $\alpha$-Te\cite{TeChiral}, where $V_{2\omega}$ displays a linear field-dependence, $\Delta R/R = \gamma^\pm \boldsymbol{B} \cdot \boldsymbol{I}$\cite{TeChiral}. 
This suggests that the magnitude of eMChA itself is field dependent, given by $\gamma^\pm(\boldsymbol{B})$.

eMChA depends on the relative direction of field and current, and hence even in a non-linear scenario, $V_{2\omega} (\boldsymbol{B})$ must change sign when the field polarity reverses, as observed experimentally. This anti-symmetric field dependence provides strong evidence against a putative thermal origin of second-harmonic voltage generation by Joule heating, as the linear magnetoresistance is even in magnetic field (see Methods).  Pronounced quantum oscillations are also observed above $B = 10$ T demonstrating an influence of Landau quantization on eMChA. This behavior is observed consistently in two devices with different mechanical mounting approaches, rendering potential torque artefacts due to the soft-mounted structure unlikely. An identically shaped sample probing in-plane transport does not show second harmonic generation at any field configuration, demonstrating eMChA to be relevant only in the interplane transport (see Methods).

To further characterize eMChA and elucidate its origin in this nearly centro-symmetric material, we next turn to the temperature dependence of $V_{2\omega}$. Figure~\ref{Cs135}b displays the raw $V_{2\omega} (\boldsymbol{B})$ without anti-symmetrization. Yet at elevated temperatures the weak thermal second-harmonic generation can obscure the chiral transport signatures, therefore we focus on the anti-symmetric component $\Delta V_{2\omega} = V_{2\omega}(18\,\mathrm{T})-V_{2\omega}(-18\,\mathrm{T})$ (see Methods for full data). At high temperatures above $T_{\mathrm{CDW}}$, no $\Delta V_{2\omega}$ is observed as expected. The transition into the CDW state is clearly evident as a sharp spike in $\Delta V_{2\omega}$ at $T_{\mathrm{CDW}}$ which we associated with the non-analyticity of $R$~($T$). A continuous anti-symmetric second harmonic signal only emerges at temperatures below 70 K. Its slow increase upon decreasing temperature suddenly accelerates at $T' \approx $ 35 K, apparent as a change in slope on the logarithmic scale. At lower temperatures, $\Delta V_{2\omega}$ increases significantly and saturates at its maximum value below 3 K. While our observations only evidence the absence of chiral scattering and do not exclude a chiral order at $T_{\mathrm{CDW}}$ that merely does not affect transport, our results are suggestive of a secondary transition or crossover at lower temperatures of $T'$. In particular, this temperature dependence agrees well with both the Fourier transformation (FT) intensity of the 4a$_0$ CDW vector (q$_{4a0}$) obtained from STM experiments\cite{zhao2021cascade} and the large anomalous Nernst effect\cite{SAxy}. Such correspondence demonstrates the direct connection between the unidirectional charge order, electronic chirality and hidden magnetic flux. This consistency is further supported by the results of muon-relaxation experiments, which suggest the onset of TRS breaking around 70~K and a subsequent rearrangement of local field distribution at 30~K\cite{Mielke2022,yu2021evidence,Nie2022}.

\subsection*{Field-switchable electronic chirality}
\begin{figure}
	\centering
	\includegraphics[width=0.9\columnwidth]{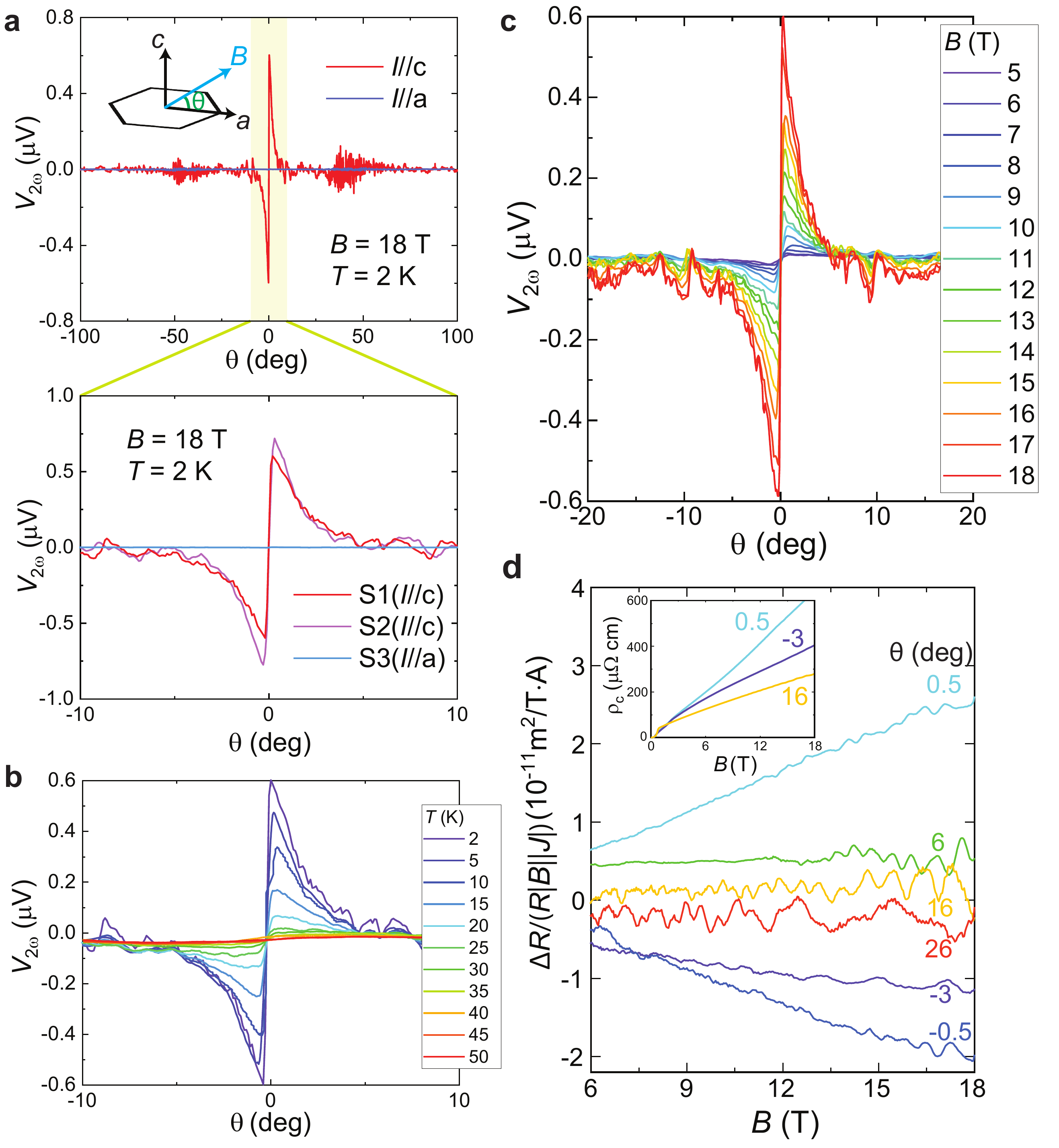}
		\caption{\textbf{Angular dependence of eMChA.} (a) Angular dependence of $V_{2\omega}$ for I$\parallel$a and I$\parallel$c. A sharp spike with sudden sign-reversal within 0.5 deg is observed for I$\parallel$c around a-axis. (b) Angular dependence of $V_{2\omega}$ at various temperatures with $B$ = 18 T and (c) of various magnetic fields at $T$ = 2 K.  (d) Field-dependent eMChA-coefficients at various angles. The inset displays the magnetoresistance measured with current applied along c-axis at various field directions. }
	\label{Main}
\end{figure}

The unusual nature of the eMChA in CsV$_3$Sb$_5$ becomes apparent when the field orientation is varied with respect to the Kagome planes ($\theta = 0^\circ$ denotes the in-plane field orientation, Fig. \ref{Main}). No $V_{2\omega}$ is observed at large field angles ($\theta > 10^\circ$). Only within a narrow angle range, $\theta=\pm 10^\circ$, $V_{2\omega}$ quickly grows as the field-angle approaches $\theta = 0^\circ$. It reaches a maximum around $\theta \sim 0.5^\circ$, the configuration discussed previously in Fig. \ref{Cs135}. For smaller $\theta$, $V_{2\omega}$ rapidly decreases and vanishes for fields within the Kagome planes ($\theta=0^\circ)$. Upon further rotation to small negative $\theta$,  the signal repeats yet with opposite sign. This marks a most striking aspect of the data: Tilting the field across the Kagome nets changes the handedness of the material. Rotating the field by $1^\circ$ barely changes $\boldsymbol{B}$, hence an abrupt sign-change of $V_{2\omega}$ implies a \textcolor{black}{transition into the opposite enantiomer}. Furthermore, the signal's magnitude strongly reduces upon raising the temperature or lowering the field strength, while the angular extent and the sharp anomaly at the in-plane field persists. At temperatures above 35~K the peak is hardly observable, and the faint residual anomaly reflects the exponential drop of $V_{2\omega}$ above $T'$ (Fig. \ref{Cs135}d). The rotation curves are slightly hysteric, however given the sharpness of the steep transition, it was impossible to distinguish a intrinsic hysteresis from the mechanical backlash of the rotator.

The possibility and ease of magnetic manipulation of the electronic chirality presents a unique electro-magnetic response of CsV$_3$Sb$_5$. It suggests that the low-temperature state differs from a simple chiral charge redistribution, as for example observed in the 3$q$ chiral CDW~\cite{CCDW} state of TiSe$_2$. Such a static charge redistribution only couples to magnetic fields via higher-order interactions, and its involved lattice response renders it unlikely to be easily manipulated at temperatures well below $T_{\mathrm{CDW}}$. Instead, the experimental situation in CsV$_3$Sb$_5$ points to coupled TRS breaking, including the concomitant magnetic anomalies at $T_{\mathrm{CDW}}$, the field-tunability, as well as muon spectroscopy experiments\cite{Mielke2022,yu2021evidence}. As a microscopic picture for this correlated state, an orbital loop current phase in the Kagome planes has been proposed that is consistent with these experimental observations\cite{TitusAdd,JPH,VhS}. 

\subsection*{Analysis of eMChA strength}
Despite its exotic properties, the eMChA in CsV$_3$Sb$_5$ can be rationalized within the existing theoretical framework. The magnetoresistance of CsV$_3$Sb$_5$ is approximately linear in $B$ for small angles $\theta$ at high magnetic fields [see inset of Fig. 3(d)]. Such a behavior is indeed not unexpected for a material with density-wave order~\cite{feng:2019} and has also been observed in many semi-metals\cite{TaAs,Na3Bi}. This marks a crucial difference to previous eMChA studies in which the conventional resistance $R(\boldsymbol{B},\boldsymbol{I}) + R(\boldsymbol{B},-\boldsymbol{I}) \approx 2 R_0$ remains approximately field-independent. This, and the strong $\theta$ dependence of the eMChA coefficient $\gamma$ means that eMChA cannot be characterized by a constant tensor, as common practice in literature on conventional eMChA materials. Yet, we can gain some insights about the magnitude of eMChA by computing 
$\Delta R~/~(R|\boldsymbol{B}||\boldsymbol{J}|) = 4V_{2\omega}~/~(V_{\omega} |\boldsymbol{B}||\boldsymbol{J}|)
$ for given magnetic field strength, see Fig.~\ref{Main}d\cite{TeChiral}, for quantitative comparisons to other systems (The quantity $\Delta R~/~(R|\boldsymbol{B}||\boldsymbol{J}|)$ equals the constant $\gamma$ when it is a field-independent parameter in chiral materials commonly used in literature.). At $B$ = 18 T and $\theta = 0.5^\circ$  we find  $\Delta R~/~ (R|\boldsymbol{B}||\boldsymbol{J}|) \approx 2.4\times$ 10$^{-11}$ m$^2$~/~T$\cdot$A. In comparison, this value is smaller than its record observations in t-Te\cite{TeChiral} (10$^{-8}$ m$^2$~/~T$\cdot$A) and TTF-ClO$_4$\cite{TTlO4} (10$^{-10}$ m$^2$~/~T$\cdot$A), where the distinct structural chirality results in relatively large eMChA, while it is larger than that of chiral magnets such as CrNb$_3$S$_6$\cite{CrNbSB}(10$^{-12}$ m$^2$~/~T$\cdot$A) and MnSi\cite{MnSi}(10$^{-13}$ m$^2$~/~T$\cdot$A), in which the chiral spin texture plays a major role in eMChA.

\begin{figure}
	\centering
	\includegraphics[width=0.97\columnwidth]{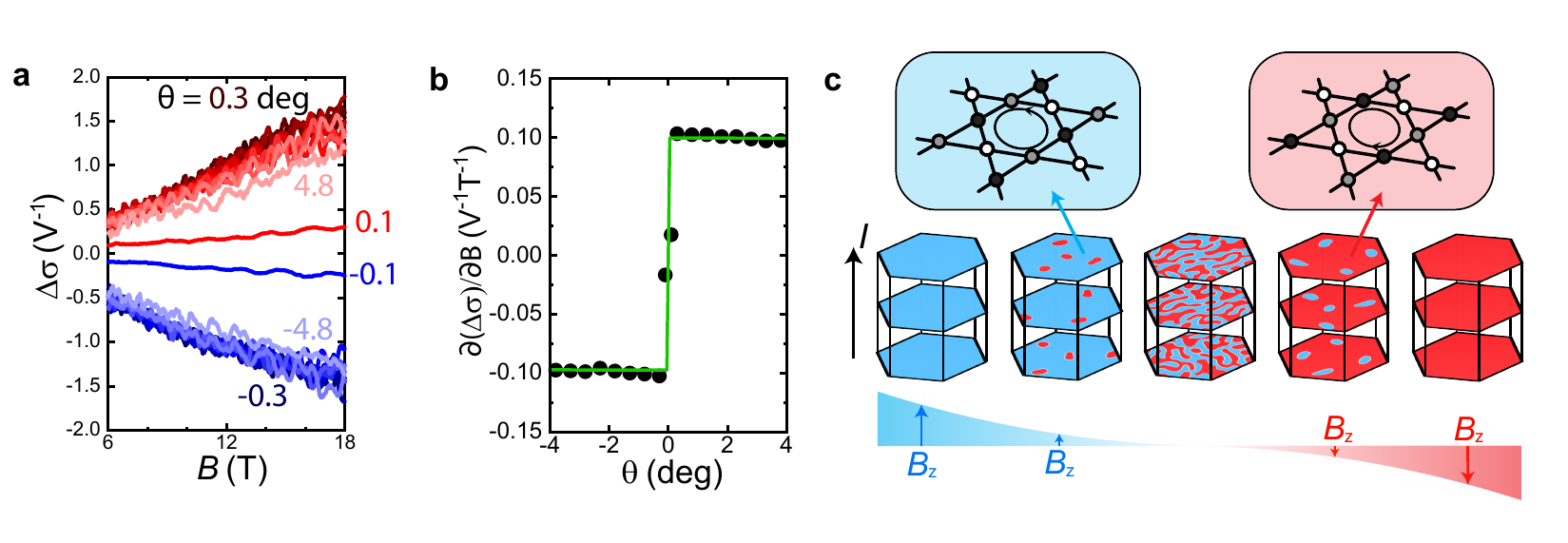}
		\caption{\textbf{Theoretical modelling of field-switchable chiral domains.} (a) Field-dependence of chiral-conductivity $\Delta \sigma$ at various angles. Between $\theta$ = $\pm$0.3  and $\pm$4.8 deg all data are measured with an angle step of 0.5 deg.  (b) Angular dependence of the first order derivative $\partial (\Delta\sigma)/\partial B$ from $B$~=~6~to~18~T. The green curve represents the model description of chiral conductivity: $ \frac{\partial \Delta \sigma}{\partial B} = {\rm sign}(\theta)\,\cos (\theta)\, \frac{\Delta_0 \tilde{M} e^2 \tau}{2 \pi^3 \hbar} $, as derived in the Methodsary Sec. IX. (c) Sketch of a chirality reversal at in-plane aligned magnetic fields. The emergence of opposite domains can naturally lead to a strong enhancement of eMChA at low field angles.}
	\label{Theory}
\end{figure}

As the conventional eMChA analysis is only applicable for materials with negligible magnetoresistance, a description in terms of the conductance is more appropriate to further capture the lowest-order field-tuned behavior of the response in CsV$_3$Sb$_5$ (see Methods).
For purely longitudinal transport and negligible Hall resistivity, the conductance is the inverse of the resistance such that in anology to the usual analysis of eMChA, we write for the conductance $\sigma + \Delta\sigma \approx 1~/~R-\Delta R~/~2R^2$ (see Methods).
We can thus extract $\Delta\sigma \propto V_{2\omega}~/~V_{\omega}^2$, where now $V_{\omega}$ is linear in $B$ for large fields. For a field applied approximately in-plane, $\Delta\sigma$ is thus approximately linear in $B$, which is the lowest-order coupling between magnetic field and current (see Methods) and naturally explains the $B^3$-dependence of $V_{2\omega}$. The linear field dependence of $\Delta\sigma$ yields a field-independent first order derivative $\partial (\Delta\sigma)/\partial B$, see Fig.~\ref{Theory}. The sudden sign reversal of $\partial (\Delta\sigma)/\partial B$ for small $\theta$ then suggests that the out-of-plane component of the field, $B_z$, has a non-perturbative effect on the system and we treat it separately, while the in-plane component is a perturbation to linear order. In other words, we write $\Delta\sigma(\boldsymbol{B}, I_z) = \tilde{\sigma}(B_z) B_x I_z$. Note that such a coupling is only allowed for a system which breaks the $y\mapsto -y$ mirror symmetry. With $\Delta\sigma(\boldsymbol{B}, I_x)$ vanishingly small, no similar conclusion can be drawn for the mirror symmetry $z\mapsto -z$.

\subsection*{Theoretical modelling}
The behavior of $\Delta \sigma$ seen in our experiment demonstrates that the charge order in CsV$_3$Sb$_5$  (i) breaks in-plane mirror symmetries, at least below $T'\sim$ 35 K and (ii) can be manipulated by an out-of plane magnetic field in the same temperature regime. We thus establish that the tunability of the chirality of charge order in CsV$_3$Sb$_5$, previously seen in STM experiments, is a macroscopic bulk property of the unconventional charge order.

We further propose the following qualitative scenario that would be consistent with the full $\theta$ dependence of our experimental observations and calls for confirmation by local-probe techniques (Fig.~\ref{Theory}). $B_z$ is the natural tuning parameter in Kagome-net physics, as evidenced by our as well as STM experiments. Akin to a soft ferromagnet, large values of $B_z$ (large $\theta$) induce a fully polarized, mono-chiral state. In this polarized state, only intrinsic chiral scattering processes induce eMChA, which commonly are weak. As the field is tilted towards the plane, $B_z$ is reduced and domains of opposite chirality appear, which act as ideal chiral scattering centers. Hence, domain wall scattering leads to strong extrinsic eMChA. Naturally, a local probe like STM would observe a chiral structure, and occasionally the required domain boundaries between them, as indeed is the experimental situation\cite{li2022rotation,Kchiral}. At even smaller $B_z$ for fields very close to the planes, both chiralities appear symmetrically and hence a globally averaging probe like transport observes a macroscopically symmetric conductor with vanishing eMChA. A fully symmetric process appears if the field is turned further, yet with inverted roles of majority and minority chirality.

In this scenario, the chirality switching is driven by $B_z$ independent of the in-plane field, in particular it would also occur for fully out-of-plane fields, where no eMChA is observed in our experiment. Yet unlike structurally chiral systems, here the magnetic field plays a dual role. While $B_z$ sensitively changes the sign of $\partial (\Delta\sigma)/\partial B$, the large in-plane field is essential to observe finite eMChA, as $\Delta\sigma\propto I_z|\boldsymbol{B}|$. Given the close relation between the field-switchable chiral transport and the chiral domains in CsV$_3$Sb$_5$, it is worth to explore its generality in other materials with suspected chiral orbital loop current.

\subsection*{Outlook}
While the small magnitude and extreme environmental conditions likely preclude direct applications of CsV$_3$Sb$_5$, it showcases that spontaneous symmetry breaking can be utilized to transform small changes in external fields into singular changes in the response functions of chiral conductors. Given the subtle deviation from centro-symmetry of the charge-ordered phase, the emergence of eMChA in correlated states calls for new theoretical approaches to identify the microscopic mechanisms. The multitude of competing ground states in correlated materials give rise to their versatility and tunability, which now presents a new approach towards chiral transport. In this direction, the field-switchable chiral transport adds a new aspect to the emergent picture of a highly frustrated, strongly interacting electron system on the Kagome planes of CsV$_3$Sb$_5$. While the magnitude of eMChA is unexpectedly large, these results well connect to recent works that have shown the charge-ordered state to be chiral and TRS breaking \cite{Kchiral,Mielke2022,yu2021evidence,Nie2022}. Akin to the theme of coupled orders in multi-ferroics, a series of new response functions emerges in materials like CsV$_3$Sb$_5$ with multiple intertwined order parameters. 

\section*{References}

\clearpage

\clearpage
\section*{Methods}
\subsection*{Crystal synthesis and characterization}

 CsV$_3$Sb$_5$ crystallizes in the P6/mmm space group which features a layered structure of Kagome planes formed by the V-atoms (Fig. S1). The single crystals were grown by the self-flux method\cite{ortiz2020cs}. Hexagonal plate-shaped crystals with typical dimensions of 2 × 2 × 0.04 mm$^3$ were obtained. The crystals were characterized by the x-ray diffraction (XRD) off the maximum surface on a PANalytical diffractometer with Cu K$\alpha$ radiation at room temperature. As shown in Extended Data Fig. 1, all the peaks in the XRD pattern can be identified as the (00$l$) reflections of CsV$_3$Sb$_5$. 
 
 Based on the crystalline structure we have calculated the band structure of CsV$_3$Sb$_5$ by density functional theory (DFT) using the Quantum Espresso package (QE)\cite{QE-2017}, the details of which can be found in Ref.~\onlinecite{huang2022mixed}. The obtained electronic structure features multiple Dirac nodal lines lifted by spin-orbit coupling (SOC), leaving only symmetry-protected Dirac nodes at L-points. These results are consistent with the previous reports \cite{ortiz2019new,ortiz2020cs}.
 
 Magnetoresistivity measurements were performed with electric current and magentic field applied along out-of-plane (z) and in-plane(x) directions respectively. The magnetoresistance displays a quasi-linear field-dependence up to $B$ = 18 T while Hall resistivity is almost negligible compared to that. This is expected as the electrical current is applied along the out-of-plane direction, Hall resistivity should vanish for such a quasi-2D material with the Brillouin zone dominated by the large cylindrical Fermi surfaces. 

\subsection*{Angular dependence of magnetoresistance and its relation to $V_{2\omega}$}
At low temperature, angle-dependent magneto-resistivity displays a strong peak when magnetic field direction rotates across the Kagome-plane (Extended Data Fig. 2). With increasing temperature this peak gradually transits to a broad hump at $T$ = 50 K. This strong enhancement of magneto-resistivity at in-plane fields is likely a feature of open-orbit magneto-transport which is expected for a metal featuring cylindrical Fermi surface sheets\cite{OpenOrbit}. The smearing of the spike at high temperatures therefore demonstrates the reduction of carrier mobility with increasing temperature. 

In the meantime, the strong increase of magneto-resistivity naturally leads to the enhancement of $V_{2\omega}$. Combining the angular dependence of both leads to a clear quadratic relation between $V_{2\omega}$ and $\rho_c$ which provides further evidence for the chiral conductance analysis.

\subsection*{IV-characteristics of eMChA}
Second harmonic voltage generation due to eMChA is expected to display a quadratic current dependence. Here we present the IV-characteristics of both first and second harmonic voltage measured with a 7~Hz AC current (Extended Data Fig. 3). For $V_{\omega}$ it depends linearly on current which corresponds to the first-order resistance term. On the other hand, the second harmonic voltage shows a clear quadratic current dependence which is an expected signature of eMChA. These results again demonstrate the observed $V_{2\omega}$ indeed originates from the chiral correction of conductivity due to electronic chirality of CsV$_3$Sb$_5$. 

\subsection*{Examination of Joule heating effect}
Joule heating is a natural extrinsic origin of higher harmonic voltage generation\cite{lu20013omega,huang3omega}. Applying an AC electric current $I_\omega$ must result in an oscillating temperature with a frequency of $2\omega$. Therefore if the electrodes of the device are strongly imbalanced in contact resistance, $V_{2\omega}$ is observed. To further check the influence of Joule heating in the measurements of eMChA in CsV$_3$Sb$_5$, we have performed systematic current-dependent $V_{2\omega}$ measurements under different thermal conditions (Extended Data Fig. 4). By controlling the helium gas pressure of the sample space, the thermal link between the device and sample chamber can be easily tuned. Within a low current regime (below 0.12 mA), the collapse of all curves measured at different conditions suggest the insignificance of Joule heating effect. With further increasing current, Joule heating inevitably grows and becomes detectable. To avoid any disturbance due to Joule heating, all measurements of eMChA  have been performed with a relatively low AC current of 0.1 mA and high gas pressure (p$_s$ $\approx$ 600 mbar) of the sample space, providing the maximal cooling power.

\subsection*{Reproducibility of eMChA with two different devices}
To show the reproducibility of the second harmonic voltage generation due to eMChA in CsV$_3$Sb$_5$, we have measured two membrane-based devices with different mounting techniques/geometries (Extended Data Fig. 5). For device S1, the sample is completely suspended by soft Au-coated membrane springs. In comparison, device S2 is attached to the membrane only on one side, the other side of the sample is welded directly to the Si-substrate by FIB-assisted Pt-deposition.
Device S2 displays a slightly broader CDW transition than S1 in the temperature dependence of resistivity across $T_{CDW}$ yet the transition temperatures are exactly the same. This suggests a marginally larger strain gradient across the device due to thermal contraction for device S2 which is compatible with the estimated strain value presented in Sec. III. The second harmonic voltage is consistently observed among the two devices with similar value as well as almost identical angular spectrum. These results demonstrate the clear consistency among different low-strain samples and therefore evidence that the observed eMChA in CsV$_3$Sb$_5$ is an intrinsic material property. These devices further differ in their coupling strength between the substrate and the device. The similarity of the data speaks in light of the much stiffer coupling in S2 speaks against magnetic torque induced angle changes as a putative error source.

\subsection*{Estimation of strain due to differential thermal contraction}
To obtain the tensile strain applied to the sample, we need to firstly estimate the total displacement due to different thermal contraction coefficients of samples and substrates used. Upon cooling from 300~K to 4~K, the integrated thermal contraction coefficient of SiN$_{x}$($\upvarepsilon_{SiN}$) and Si($\upvarepsilon_{Si}$) are 0.0342\% and 0.0208\% respectively. For the sample itself we assume a typical thermal contraction coefficient for alkali metal $\upvarepsilon_{Samp}$ $\approx$ 0.1\%, which provides a conservative, upper bound. Based on these parameters and the actual device geometry illustrated in Extended Data Fig. 6, the total displacement can be easily obtained as:
\begin{align}
dL_{S1}=L_{S1} \times \upvarepsilon_{Samp} = 30~nm&
\\dL_{SiNx}=L_{SiNx} \times \upvarepsilon_{SiNx} = 75~nm&
\\dL_{Si}=L_{Si} \times \upvarepsilon_{Si} = 52~nm&
\\dL_{S1}=dL_{S1}+dL_{SiNx}-dL_{Si} = 53~nm&
 \end{align}

The spring constant of the SiN$_x$ spring for device S1 is estimated as 125 N/m from finite element simulations\cite{huang2022mixed} (COMSOL), the total pressure can be calculated as: 

\begin{equation}
P_{S1}=k_{S1} \cdot dL_{S1} / A = 8.8~bar
 \end{equation}

where A stands for the cross section of the spring.

Meanwhile for device S2, the pressure can be calculated with the same process:
\begin{equation}
P_{S2}=k_{S2} \cdot dL_{S2} / A = 18.7~bar
\end{equation}
In both cases the pressure is less than 20~bar. Taking the typical Young's modulus of alkali metals ($\sim$ 5 GPa), the strain applied on the sample is estimated to be $\sim$ 0.04\%, which quantifies the low-strain nature of these devices.

\subsection*{Strain effect on eMChA}
The necessity of low-strain mounting is revealed by a comparative study of device S4 which features a sample that is glued down to a sapphire substrate. Here the device is structured into an "L"-shape with two long beams along both $a$ and $c$-direction respectively (Extended Data Fig. 7). Since the sample and substrate are mechanically coupled via the glue droplet, the thermal contraction difference between them results in a tensile strain along the beam direction. This tensile strain not only shifts the CDW transition of device S4 to a higher temperature compared to the strain-free S1, but also suppresses the superconducting transition down to lower temperature. Most importantly, no meaningful second harmonic voltage has been observed for device S4. These observations suggest the importance of c-axis tensile strain which in defining an extrinsic, long-range domain structure - unable to be switched or tuned. This observation suggests that residual strain fields would provide a natural explanation for the contradictory Scanning tunneling microscope(STM) experimental results\cite{Kchiral,li2022rotation}.

\subsection*{Field-symmetry analysis of second harmonic voltage}
To further demonstrate the origin of second harmonic voltage generation, we also measured the temperature-dependent $V_{2\omega}$ at $B$ =  18 and -18 T respectively (Extended Data Fig. 8). By taking the sum and difference of these two results we obtained both the field-symmetric and asymmetric components of $V_{2\omega}$. It is clear that the anti-symmetric component dominates the total signal at low temperatures, while the symmetric component most likely due to Joule heating at the electric contacts is merely a minor part.  

\subsection*{Theoretical Considerations}
In this section, we discuss the magnetoresistance of a single-band model to illustrate the appearance of the various contributions to the linear magnetoresistence and to the second order response discussed in the main text. In particular, we are interested in the effect of a charge-density wave (CDW) on the magnetotransport, when the CDW not only breaks translation, but also time-reversal and several mirror symmetries. Note that we focus here on intrinsic contributions which allow explain the abrupt switching of the second order response at small $\theta$. To model the full $\theta$ dependence, extrinsic contributions would have to be included in addition, as discussed in the main text.

In most metals, the transverse magnetoresistance scales quadratically with magnetic field, $\rho_{zz}(B_x) \propto B_x^2$. However, this behavior can change to $B$ linear if there are small Fermi surfaces or Fermi surfaces with sharp corners~\cite{abrikosov:2000}. While the conditions for such $B$-linear behavior are most probably not satisfied in the normal state of CsV$_3$Sb$_5$, the Fermi surface reconstruction due to the CDW instability is expected to result in new, smaller Fermi surfaces, such that a linear magnetoresistance as observed can be explained. Note that linear magnetoresistance in density-wave materials has indeed been observed and discussed in the context of Fermi-surface reconstruction by Feng and coworkers~\cite{feng:2019}. 

When the density-wave instability breaks additional symmetries, we find further contributions to the magnetoresistance, or equivalently the conductivity. To see this, we employ Boltzmann transport theory in the relaxation-time approximation, where the conductivity is given by
\begin{equation}
	\sigma_{ij} = \frac{e^2 \tau}{4\pi^3\hbar^2} \int d^3k v_i(\bk) v_j(\bk)\frac{\partial f(\xi)}{\partial\xi}
	\label{eq:botzmann}
\end{equation}
with $e$ the electron charge, $\tau$ the scattering time, and
\begin{equation}
	\hbar \boldsymbol{v}(\bk) = \nabla_{\bk} \xi_{\bk}
	\label{eq:vel}
\end{equation}
the velocity for electrons with dispersion $\xi_{\bk}$. We assume in the following that we are in the symmetry-broken charge-ordered phase and the dispersion is given by $\xi_{\bk}^{\rm CDW}$ at zero magnetic field. If we minimally couple the vector potential to the momentum, we first find the dominant term $\sigma_{zz}^0\propto1/|B|$ as discussed above.

In addition to minimal coupling, a magnetic field can couple to the electron dispersion directly, if its Bloch states have an orbital magnetic moment $\boldsymbol{M}(\bk)$. In this case, we can write
\begin{equation}
	\xi_{\bk} = \xi_{\bk}^{\rm CDW} + \boldsymbol{M}(\bk)\cdot\bB.
	\label{eq:fulldispersion}
\end{equation}
Note that $\boldsymbol{M}(\bk)$ is a pseudovector. If TRS is present, $\boldsymbol{M}(\bk)$ is an odd function of $\bk$. If in addition inversion symmetry is present, we find $\boldsymbol{M}(\bk)\equiv 0$. This should be the case above the CDW transition temperature. In the CDW phase, in contrast, the various broken symmetries allow for $\boldsymbol{M}(\bk)$ to be nonvanishing. For concreteness, we will now only discuss the case of broken $x$-mirror symmetry, which is sufficient to explain the experimentally observed response. In this case, we find to lowest order in $\bk$ that $M_x(\bk) \propto k_z$. If, further, this symmetry breaking is due to an ordered phase with order parameter $\Delta^{\rm CDW}$, we can expand in both $\bk$ and $\Delta^{\rm CDW} $ and write $M_x(\bk) \approx \tilde{M} \Delta^{\rm CDW} k_z$ with $\tilde{M}$ a model-dependent constant. 

We can now use Eqs.~\eqref{eq:vel} and \eqref{eq:fulldispersion} to calculate the velocity in the $z$ direction,
\begin{equation}
    \hbar v_z(\bk) = v_z^0(\bk) + \frac{\partial\boldsymbol{M}(\bk)}{\partial k_z}\cdot \bB \approx v_z^0(\bk) + \Delta^{\rm CDW} \tilde{M} B_x.
\end{equation}
(More generally, $M_x(\bm{k})$ will be an odd function of $k_z$, such that the additional contribution to the velocity will be even.) Finally, we find for the conductivity in $z$ direction (dropping for simplicity the subscripts $zz$)
\begin{eqnarray}
    \sigma(\bm{B}, \bm{I}) &=& \sigma + \frac{\Delta^{\rm CDW} \tilde{M} B_x e^2 \tau}{2\pi^3\hbar} \int d^3k v^0_z(\bk) \frac{\partial f(\xi)}{\partial\xi} + O\left[B (\Delta^{\rm CDW})^2\right]\\
    &\approx& \sigma + \frac{\Delta^{\rm CDW} \tilde{M} e^2 \tau}{2\pi^3\hbar} B_x I_z,
    \label{eq:magnetocond}
\end{eqnarray}
where we have used the stationary Fermi distribution to calculate the current carried by the system. We thus find an additional contribution to the conductivity $\Delta \sigma \propto B_x I_z$, which will result in a second-harmonic signal in an AC electric field applied in the $z$ direction.

To further support the above arguments for the case of CsV$_3$Sb$_5$, we have performed a tight-binding calculation that shows in the chiral CDW phase (i) the Fermi surface structure  becomes indeed more structured with sharp corners and (ii) a finite orbital magnetic moment $M_x$ arises, see Extended Data Fig. 9.

Note that experimentally, we find that a small magnetic field in the $z$ direction can change the sign of the observed signal. This implies that a magnetic field $B_z$ couples linearly to the order parameter $\Delta^{\rm CDW}$, which in turn implies that the order breaks TRS in addition to the mirrors $M_x$ and $M_y$. This is in agreement with the experimental findings in Ref.~\onlinecite{Kchiral}. In terms of a simple Landau theory,
\begin{equation}
    F[\Delta^{\rm CDW}] = \alpha (\Delta^{\rm CDW})^2 + \frac{\beta}{2} (\Delta^{\rm CDW})^4 + \gamma B_z\Delta^{\rm CDW},
\end{equation}
with $\alpha<0$, $\beta > 0$, and $\gamma \ll |\alpha|, \beta$, the CDW order parameter is $\Delta^{\rm CDW} \approx {\rm sign}(B_z) \sqrt{-\alpha / \beta} = {\rm sign}(\theta) \Delta_0$, with $B_z = |\bm{B}|\sin \theta$, implying its sign change with $B_z$.

In summary, we find for the slope of the chiral conductivity,
\begin{equation}
    \frac{\partial \Delta \sigma}{\partial B} = {\rm sign}(\theta)\,\cos (\theta)\, \frac{\Delta_0 \tilde{M} e^2 \tau}{2 \pi^3 \hbar} .
\end{equation}
For small angles $\theta$ off the basal plane, this yields a step function in good qualitative agreement with the experimentally extracted form for small angles $\theta$. Moreover, combining the angular dependence of both the theoretically predicted chiral conductivity and the experimentally measured magnetoresistance, we also derived the second harmonic voltage $V_{2\omega}$ as a function of angle (Extended Data Fig. 10) which is also consistent with the experimental results.  

Finally, if we want to compare the conductivity calculated above to the experiment and the standard eMChA literature, we need to express the conductivity in Eq.~\eqref{eq:magnetocond} as a resistance. Namely, one usually writes $R(\bm{B}, \rm{I})=R + \Delta R$/2 where in general, $R$ can depend on $B$ and in particular, here we have $R \approx |B|$. We thus find
\begin{equation}
    \sigma(\bm{B},\bm{I}) = 1/R(\bm{B},\bm{I}) \approx 1/R - \Delta R / 2R^2.
\end{equation}
This yields $\Delta \sigma \approx - \Delta R / 2R^2$ and in turn, we expect
\begin{equation}
    \Delta R \approx 2\Delta \sigma R^2
\end{equation}

By applying a low-frequency AC current $I_\omega~=~I_0sin(\omega t)$, the generated electric voltage can be expressed as:
\begin{equation}
   \begin{split}
     V &= I_\omega \cdot (R+\Delta R/2) \\ &\approx I_{\omega}R + \tilde{\sigma} BR^2 \cdot  I_\omega^2 \\ &= I_0 R sin(\omega t) +  \frac{\tilde{\sigma} BR^2I_0^2}{2} - \frac{\tilde{\sigma} BR^2I_0^2}{2} cos(2\omega t) \\ &= V_{\omega} \cdot sin(\omega t) + V_{DC} - V_{2\omega} \cdot cos(2\omega t),
   \end{split}
   \end{equation}
   here $V_{DC}$ stands for the DC background voltage. This yields: 
   \begin{equation}
   \begin{split}
     V_{2\omega}/V_{\omega} = \frac{\tilde{\sigma}BRI_0}{2} = \frac{1}{4} \frac{2\Delta \sigma R^2}{R} = \frac{1}{4} \frac{\Delta R}{R},
   \end{split}
   \end{equation}
   which also suggests:
   \begin{equation}
   \begin{split}
     V_{2\omega} \propto B R^2 \propto B_x^3 ,
   \end{split}
   \end{equation}
   exactly in line with our experimental data.

\subsection*{Method Reference}

\clearpage

\noindent \textbf{Acknowledgements: } This work was funded by the European Research Council (ERC) under the European Union’s Horizon 2020 research and innovation programme (MiTopMat - grant agreement No. 715730 and PARATOP - grant agreement No. 757867). This project received funding by the Swiss National Science Foundation (Grants  No. PP00P2\_176789). M.G.V., I. E. and M.G.A. acknowledge the Spanish   Ministerio de Ciencia e Innovacion  (grant PID2019-109905GB-C21). M.G.V., C.F., and T.N. acknowledge support from FOR 5249 (QUAST) lead by the Deutsche Forschungsgemeinschaft (DFG, German Research Foundation). This work has been supported in part by Basque Government grant IT979-16. This work was also supported by the European Research Council Advanced Grant (No. 742068) “TOPMAT”, the Deutsche Forschungsgemeinschaft (Project-ID No. 247310070) “SFB 1143”, and the DFG through the W\"{u}rzburg-Dresden Cluster of Excellence on Complexity and Topology in Quantum Matter ct.qmat (EXC 2147, Project-ID No. 390858490).

\noindent \textbf{Author Contributions:} Crystals were synthesized and characterized by D.C. and C.F.. The experiment design, FIB microstructuring, the magnetotransport measurements and the second harmonic voltage measurements were performed by C.G., C.P., S.K., X.H. and P.J.W.M.. M.H.F. and T.N. developed and applied the general theoretical framework, and the analysis of experimental results has been done by C.G., C.P. and P.J.W.M.. Band structures were calculated by M.G.A., I.E. and M.G.V.. All authors were involved in writing the paper.

\noindent \textbf{Competing Interests:} The authors declare that they have no competing financial interests.\\

\noindent \textbf{Data Availability:} Data that support the findings of this study is deposited to Zendo with the access link: https://doi.
org/10.5281/zenodo.6787797.

\clearpage
\renewcommand{\figurename}{Extended Data Fig.}
\setcounter{figure}{0}    
\begin{figure}
	\centering
	\includegraphics[width=0.97\columnwidth]{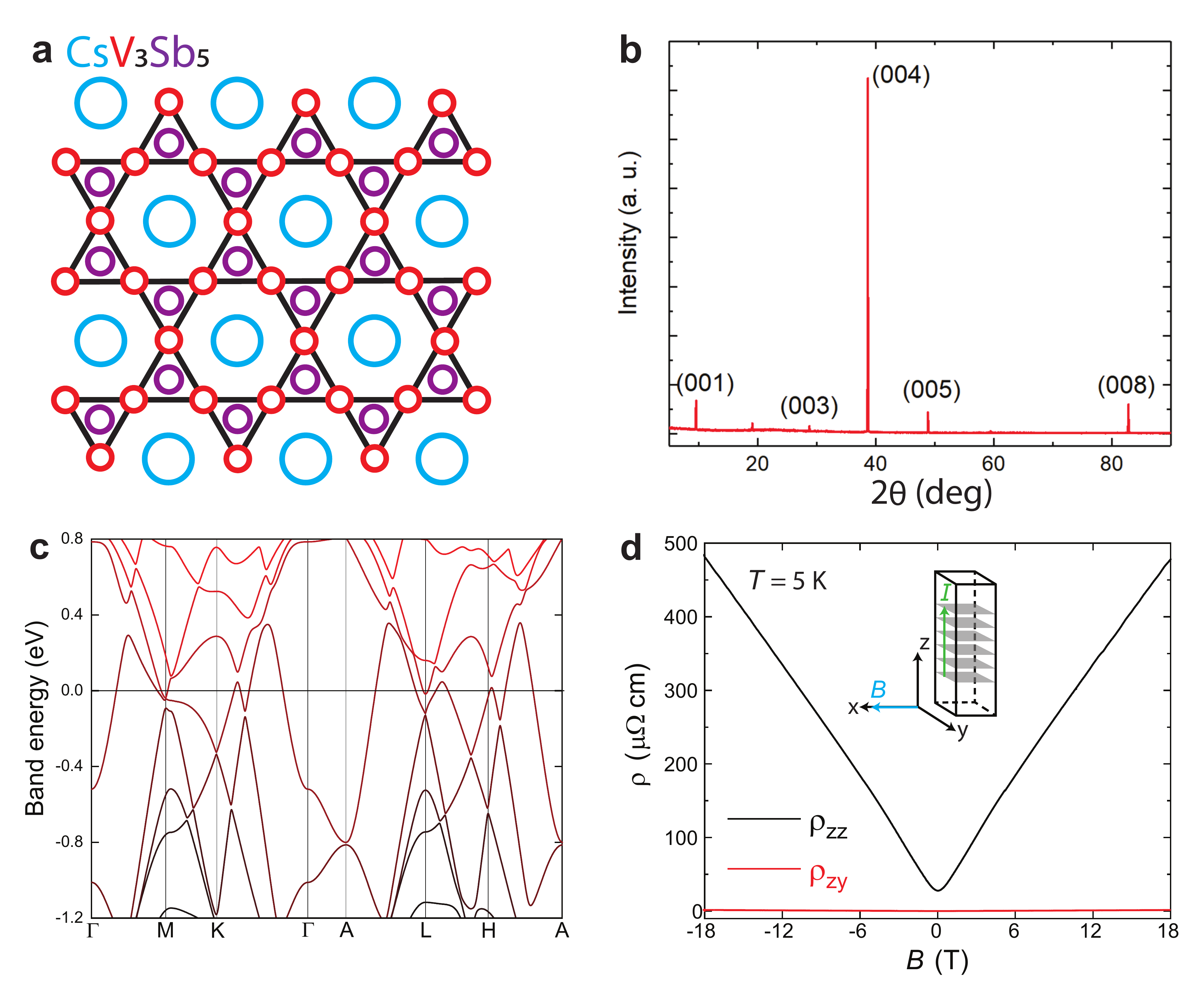}
		\caption{\textbf{Basic properties of CsV$_3$Sb$_5$.} (a) Crystal structure of CsV$_3$Sb$_5$. (b) XRD pattern of the (001) facet of a CsV$_3$Sb$_5$ crystal.  (c) Band structure of CsV$_3$Sb$_5$ calculated by density functional theory (DFT) using the Quantum Espresso package (QE)\cite{QE-2017}. (d) Field dependence of Magnetoresistivity and Hall resistivity measured at $T$ = 5 K. A large quasi-linear magnetoresistance is observed up to $B$ = 18 T. In comparison, the Hall resistivity is almost negligible.}
	\label{Basic}
\end{figure}

\begin{figure}
	\centering
	\includegraphics[width=0.97\columnwidth]{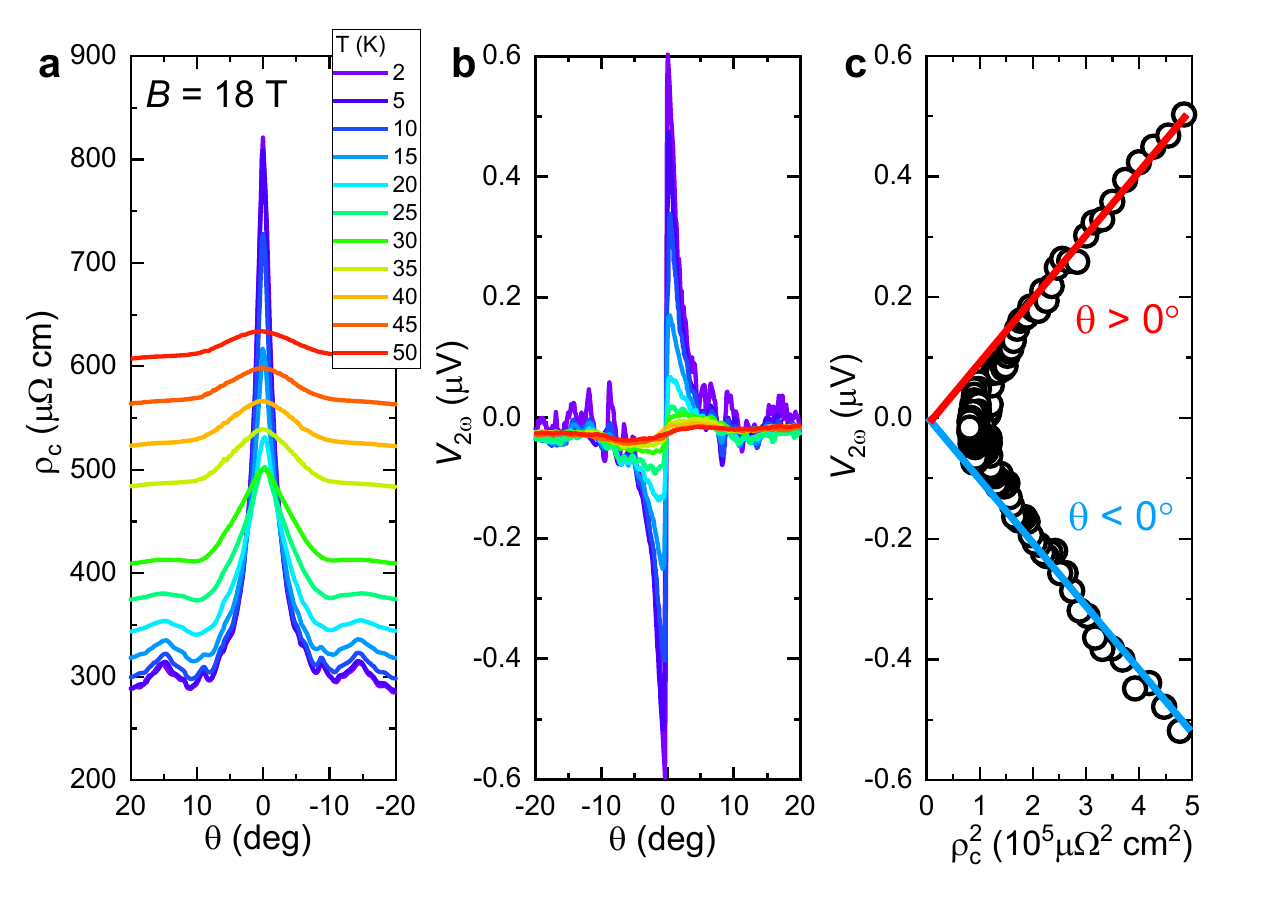}
		\caption{\textbf{Angular dependence of magnetoresistivity and eMChA.} (a) Angle-dependent magnetoresistivity of CsV$_3$Sb$_5$ measured with $B$ = 18 T at various temperatrues. A strong spike can be observed around $\theta $ = 0 deg which becomes broader with increasing temperatures. (b) Correspondingly, $V_{2\omega}$ also gets pronounced within the same angle range. (c) The $V_{2\omega}$ depends linearly on the square of c-axis resistivity, which demonstrates the direct connection between magneto-resistivity and eMChA.} 
	\label{MRangle}
\end{figure}

\begin{figure}
	\centering
	\includegraphics[width=0.97\columnwidth]{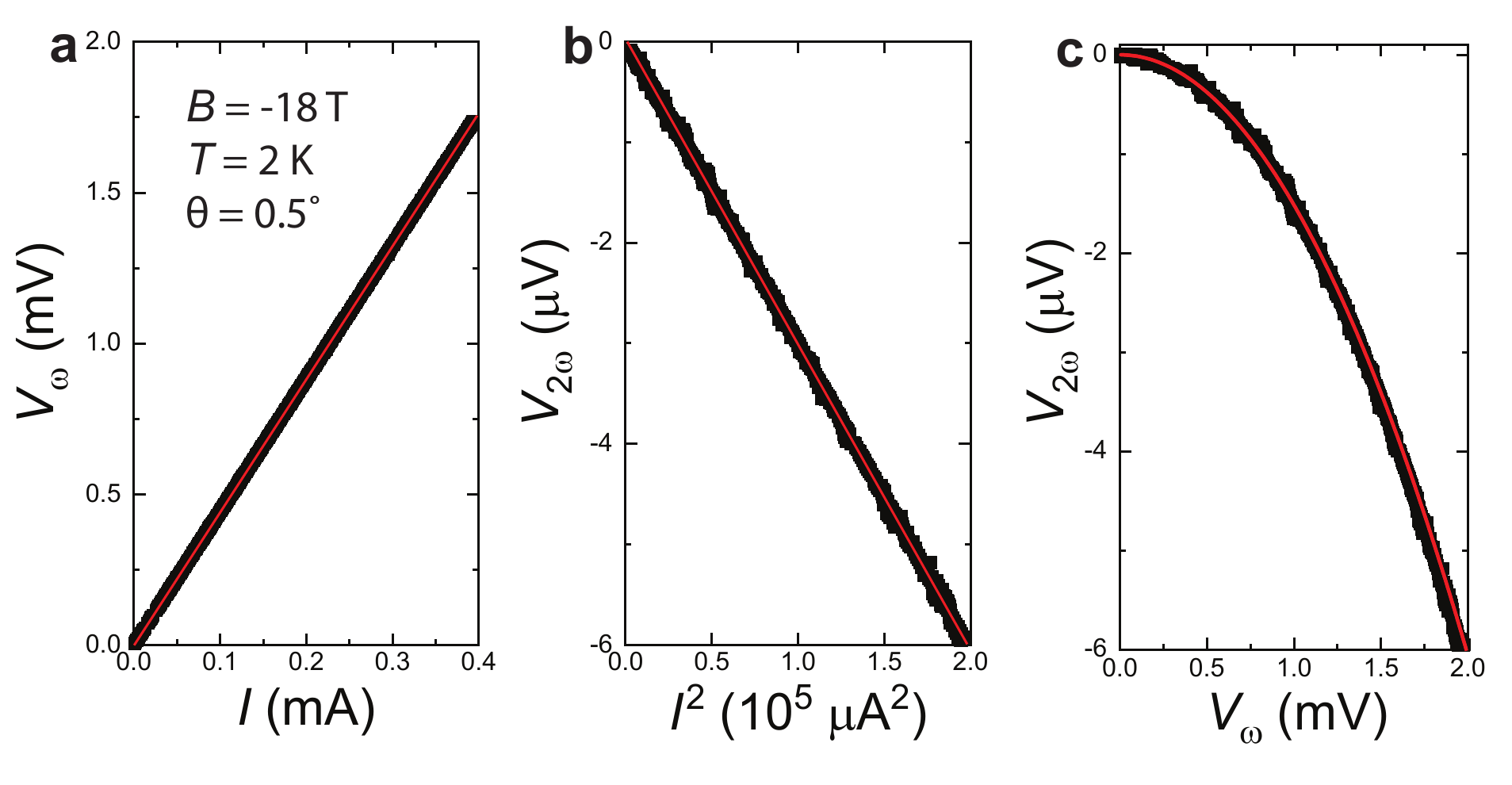}
		\caption{\textbf{Current dependence of first and second harmonic voltage.} (a) and (b) display the current-dependence of both $V_{\omega}$ and $V_{2\omega}$. As expected $V_{\omega}$ depends linearly on current while $V_{2\omega}$ displays a quadratic current dependence. (c) summarizes the relation between $V_{2\omega}$ and $V_{\omega}$, which shows a parabolic dependence.} 
	\label{IV}
\end{figure}

\begin{figure}
	\centering
	\includegraphics[width=0.97\columnwidth]{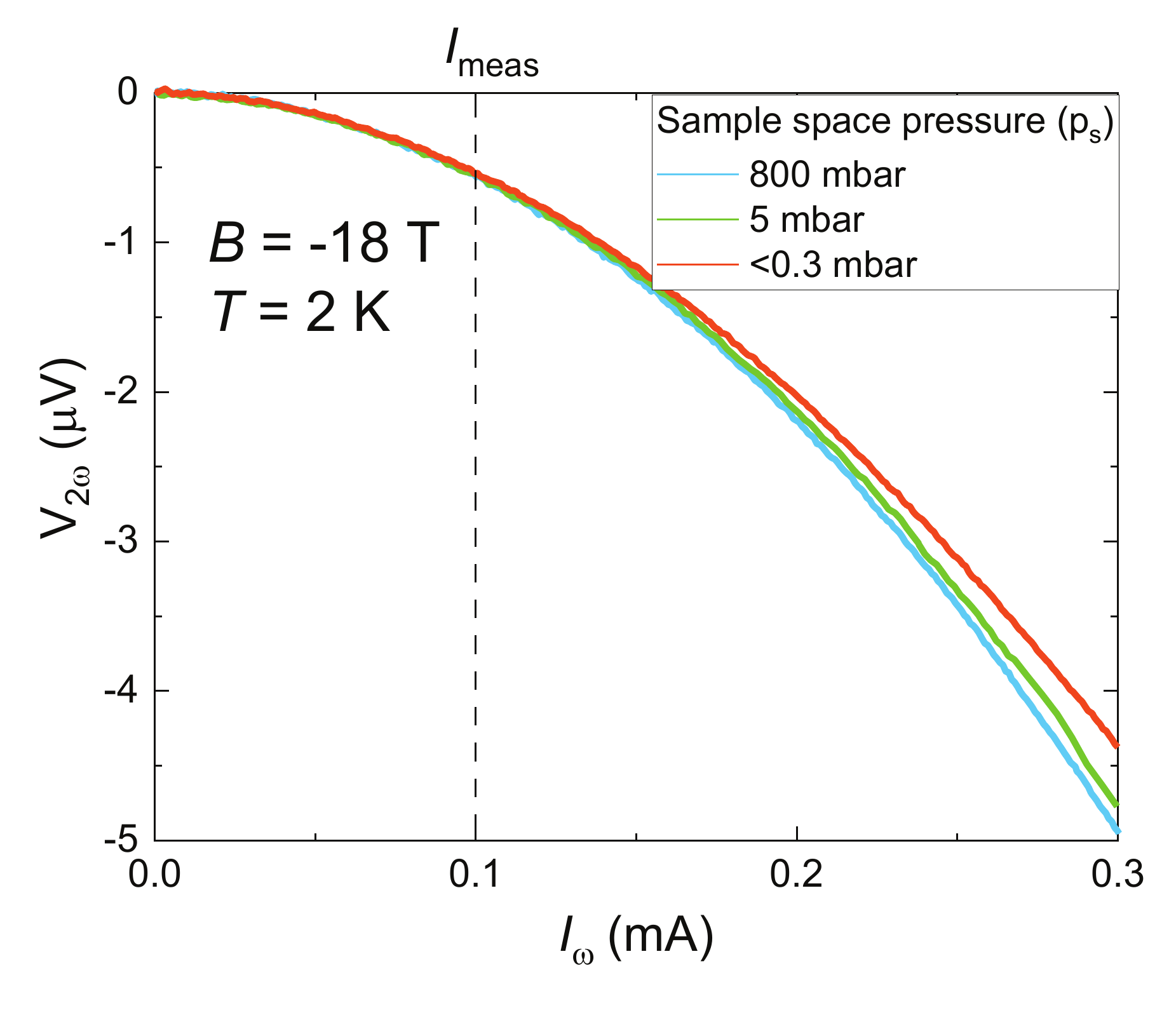}
		\caption{\textbf{Influence of Joule heating effect.} Current-dependence of $V_{2\omega}$ measured at $B$ = 18 T and $T$ = 2 K with varying levels of helium exchange gas pressure in the cryostat. The curves differ only at currents above 0.12~mA, suggesting that the heat generation and accumulation at lower currents is not a dominant factor. eMChA was measured at lower values of 0.1~mA.} 
	\label{heat}
\end{figure}

\begin{figure}
	\centering
	\includegraphics[width=0.97\columnwidth]{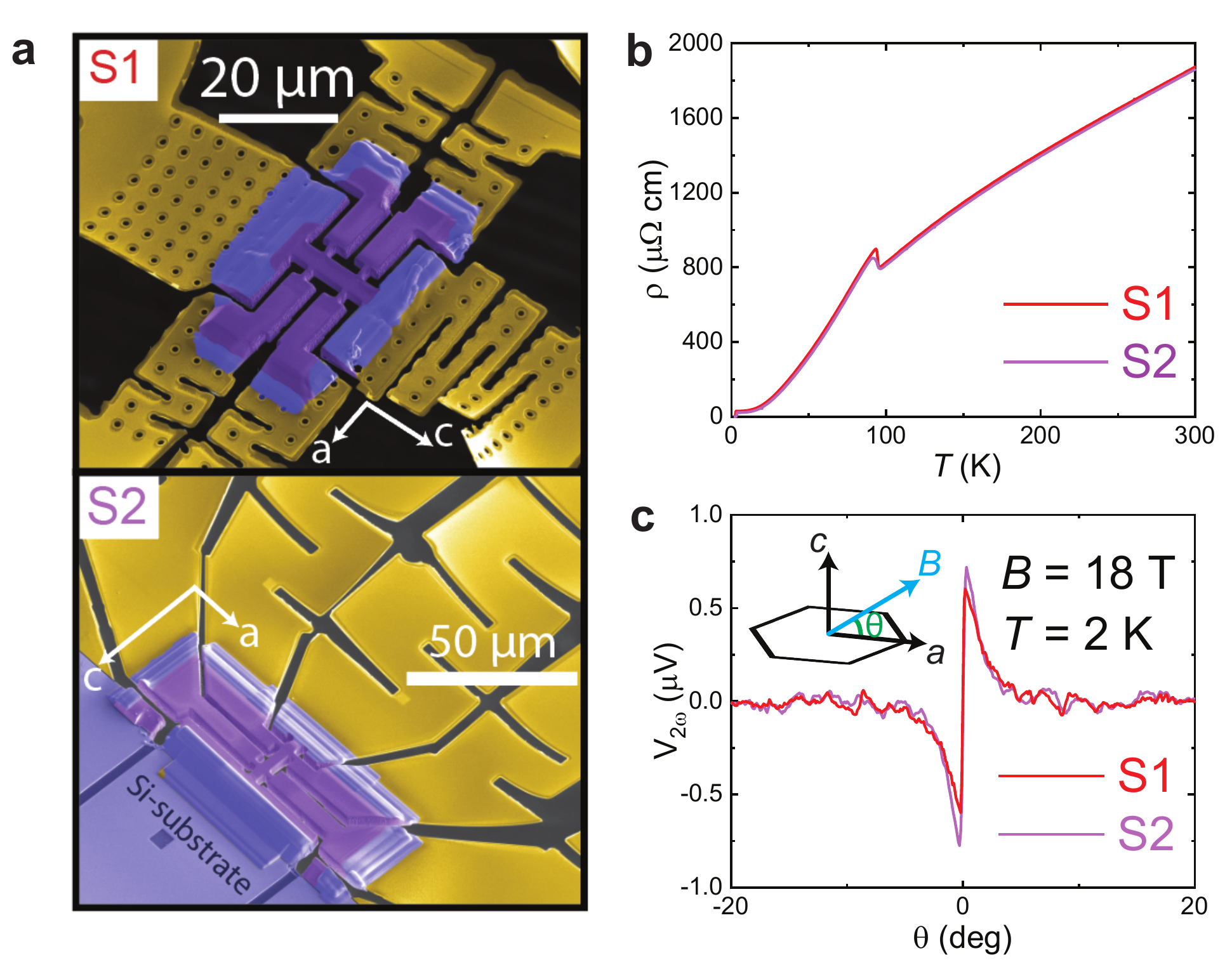}
		\caption{\textbf{Reproducibility of eMChA with two different devices.} (a) Scanning electron microscope (SEM) images of both device S1 and S2. (b) Temperature-dependent resistivity of S1 and S2 from 300~K to 1.6~K. (c) Angle-dependent second harmonic voltage measured at $B$ = 18 T and $T$ = 2 K.}
	\label{S1,2}
\end{figure}

\begin{figure}
	\centering
	\includegraphics[width=0.97\columnwidth]{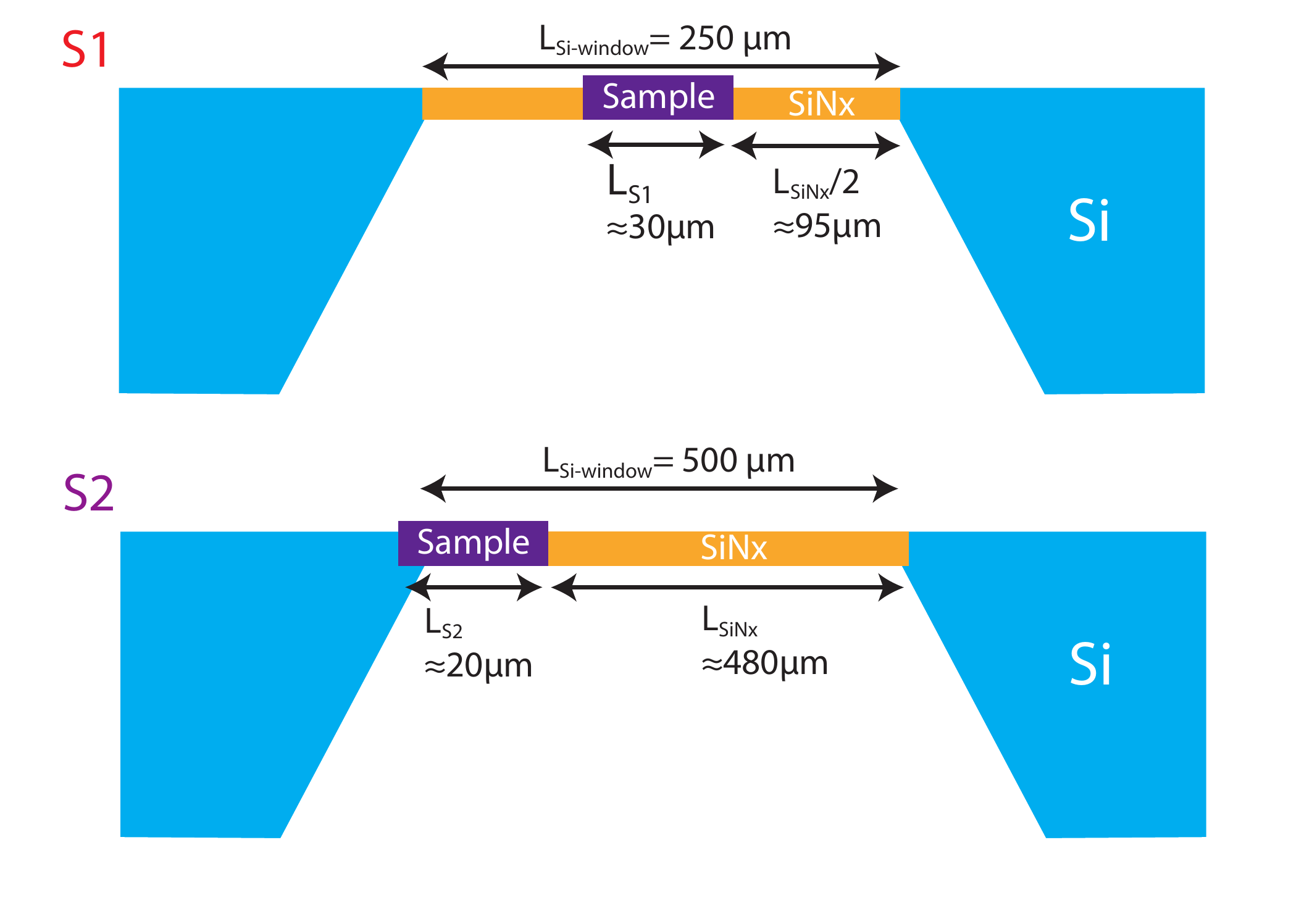}
		\caption{\textbf{Illustration of device configuration for both S1 and S2.} While S1 is completely suspended by the membrane springs, S2 is attached to the Si-substrate frame on one side and membrane springs on the other side.}
	\label{mem}
\end{figure}

\begin{figure}
	\centering
	\includegraphics[width=0.97\columnwidth]{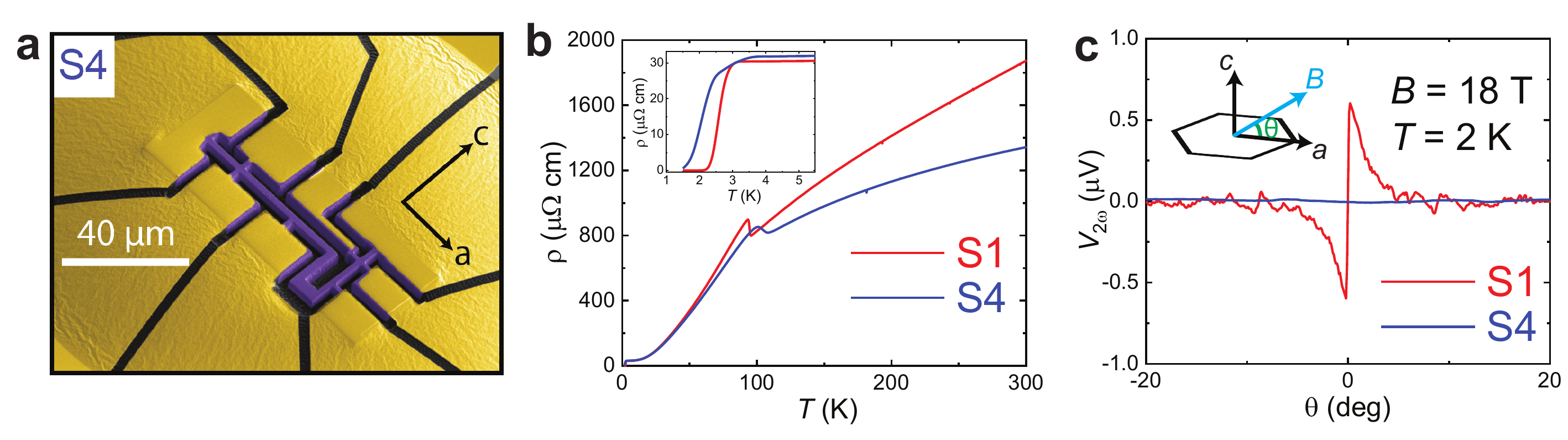}
		\caption{\textbf{Strain effect on eMChA.} (a) SEM image of device S4. The sample is attached to the sapphire substrate via a glue droplet. The thin beam along c-axis allows us to measure the electric response with both current and tensile strain applied along c-axis. (b) Temperature dependence of resistivity for S1 and S4. The CDW transition is enhanced to a higher temperature with tensile strain along c-axis (S4), while $T_c$ is reduced to lower temperature. (c)Angular spectrum of second harmonic voltage ($V_{2\omega}$). Clearly $V_{2\omega}$ is completely suppressed with tensile strain along c-axis .}
	\label{strain}
\end{figure}

\begin{figure}
	\centering
	\includegraphics[width=0.97\columnwidth]{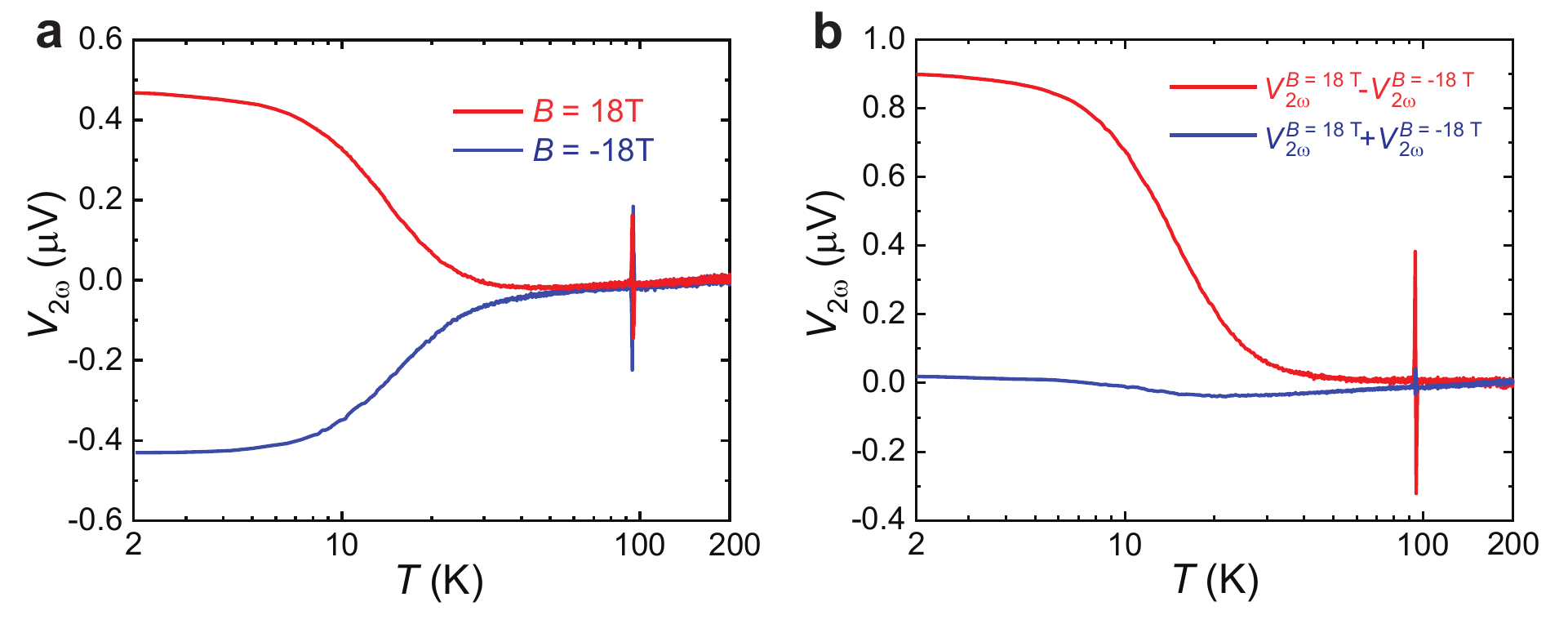}
		\caption{\textbf{Field-symmetry analysis of second harmonic voltage.} (a) Temperature dependence of $V_{2\omega}$ measured at $B$ = $\pm$ 18 T respectively. (b) Temperature-dependent field-symmetric and asymmetric part of $V_{2\omega}$.}
	\label{ASM}
\end{figure}

\begin{figure}
	\centering
	\includegraphics[width=0.97\columnwidth]{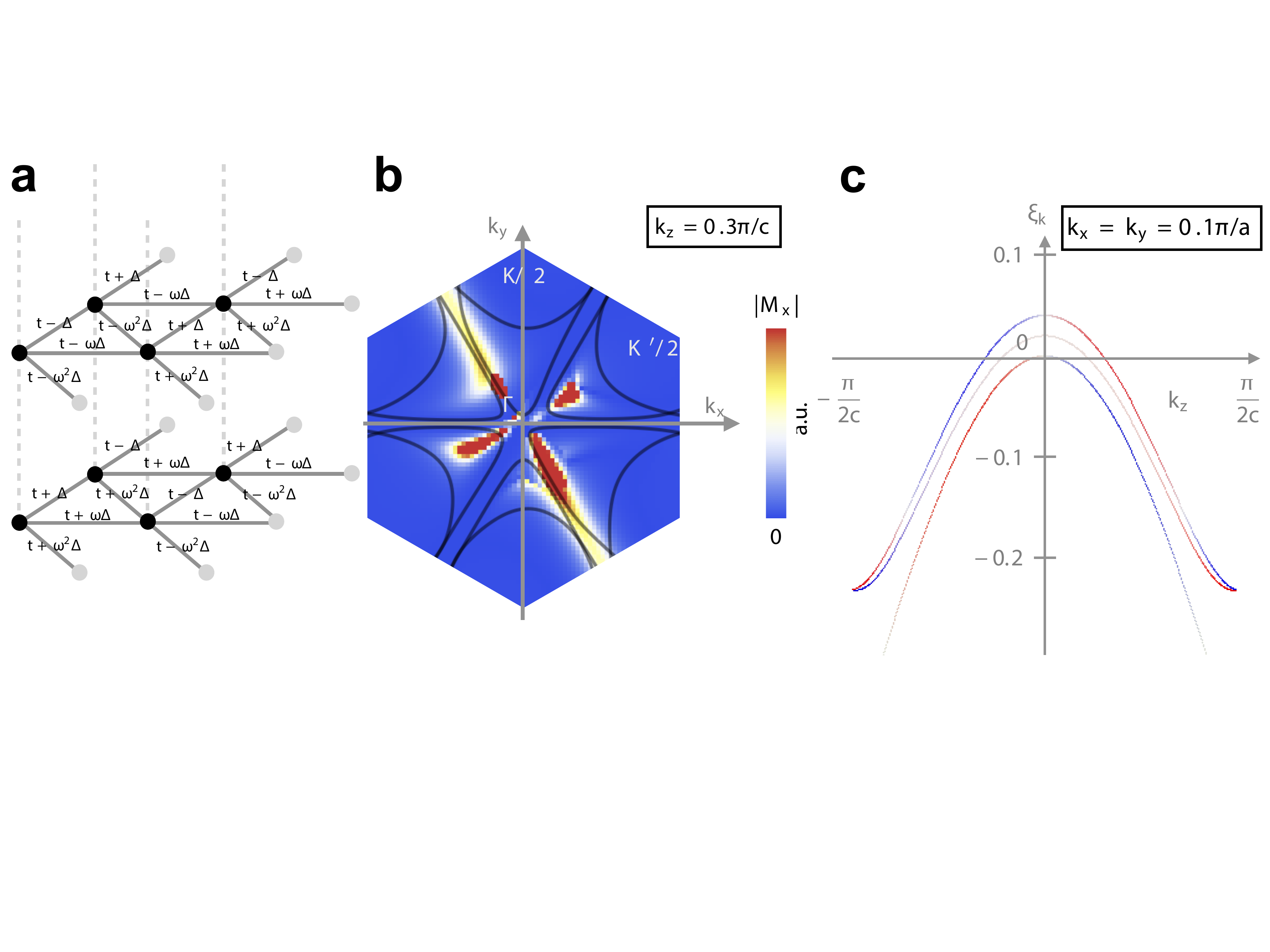}
		\caption{\textbf{Tight-binding model on the stacked triangular lattice with $2\times2\times2$ chiral CDW order.} (a) The unit cell of the model consists of 8 lattice sites (black) on two layers, each harboring a single electronic orbital. The model is defined by the hoppings as indicated, where $\omega=e^{i2\pi/3}$ (all complex hoppings are oriented to the right). All dashed lines are associated with real hopping amplitude $t_z$. We choose numerical values $t=1$, $t_z=0.2$, $\Delta=0.05$. (b) Fermi surface of the model for chemical potential $\mu=2.4$ (black lines) and orbital magnetic moment component $M_x$ of the states closest to the Fermi level. We observe that $M_x$ is not only finite, but also large for states on the Fermi surface.
		(c) Band structure along $k_z$ for a particular choice of $k_x$ and $k_y$, colored by the orbital magnetic moment component $M_x$, demonstrating that  $M_x$ is an odd function of $k_z$ as required for the measured nonlinear response to be nonvanishing.}
	\label{Model}
\end{figure}

\begin{figure}
	\centering
	\includegraphics[width=0.97\columnwidth]{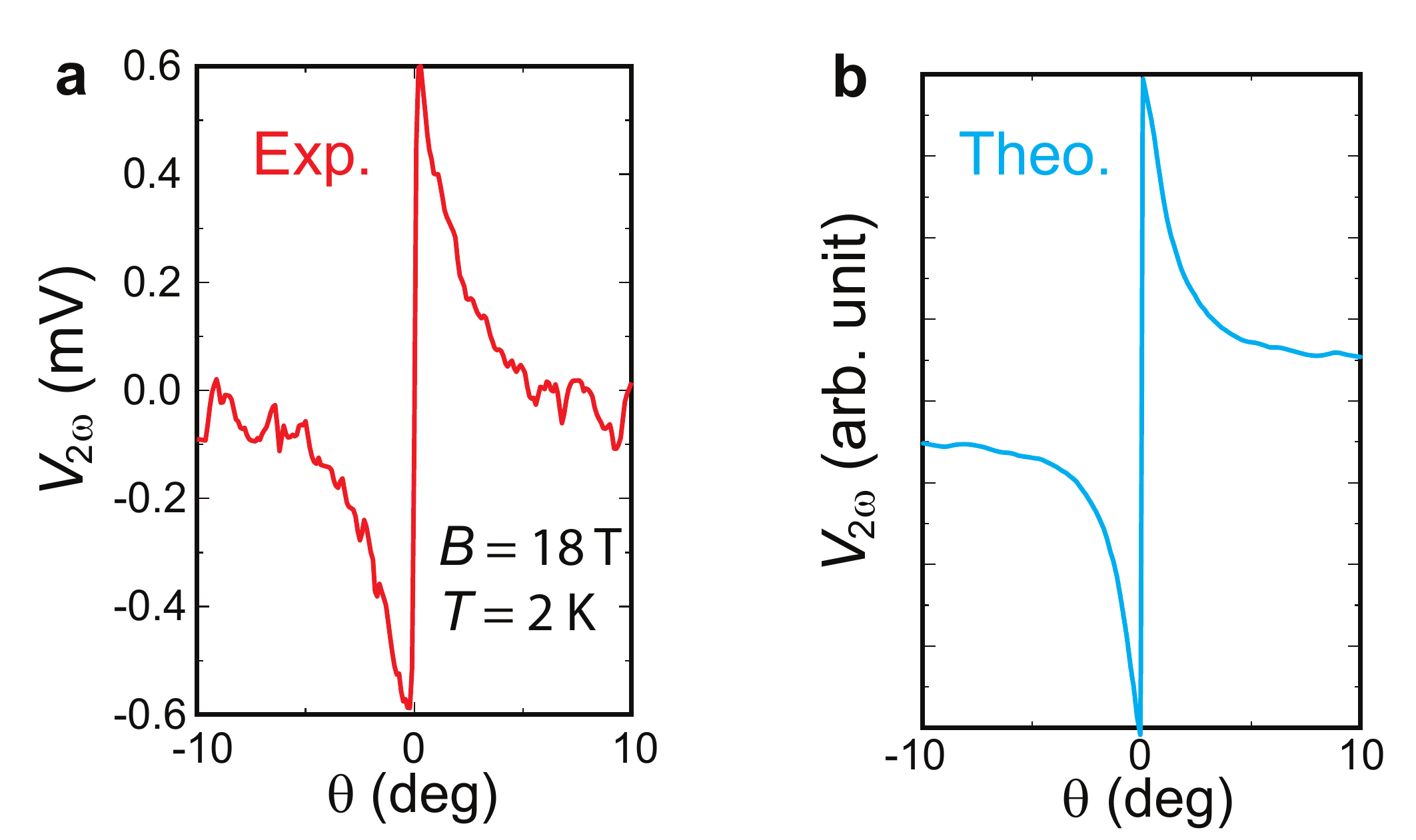}
		\caption{\textbf{Comparison between experimental results and theoretical prediction.} (a)Experimentally measured and (b) theoretically predicted angular dependence of second harmonic voltage $V_{2\omega}$ with $B$ = 18 T and $T$ = 2 K. 
		}
	\label{Fit}
\end{figure}

\end{document}